%% file: main.tex
\documentclass[conference]{IEEEtran}
\IEEEoverridecommandlockouts
\usepackage{graphicx} 
\usepackage{diagbox}
\usepackage{array}
\usepackage{float}
\usepackage{multirow}
\usepackage{hyperref}
\usepackage{longtable}
\usepackage{booktabs}
\usepackage{siunitx}
\usepackage{changepage}
\usepackage{balance}
\usepackage{pdflscape}
\usepackage{soul}
\usepackage[normalem]{ulem}
\usepackage{multicol}
\usepackage{placeins}
\usepackage{caption}
\PassOptionsToPackage{hyphens}{url}\usepackage{hyperref}
\usepackage{xcolor}  
\usepackage{todonotes}
\usepackage{adjustbox}
\usepackage{tikz}
\usetikzlibrary{positioning,shapes.misc,arrows.meta,calc}

\def\BibTeX{{\rm B\kern-.05em{\sc i\kern-.025em b}\kern-.08em
    T\kern-.1667em\lower.7ex\hbox{E}\kern-.125emX}}

\usepackage{colortbl}
\definecolor{darkgreen}{rgb}{0.0, 0.5, 0.0}

\renewcommand{\abstractname}{ABSTRACT}

\title{Prompting the Priorities: A First Look at Evaluating LLMs for Vulnerability Triage and Prioritization} 
\author{
\IEEEauthorblockN{Osama Al Haddad\IEEEauthorrefmark{1},
Muhammad Ikram\IEEEauthorrefmark{1},
Ejaz Ahmed\IEEEauthorrefmark{2}, and
Young Lee\IEEEauthorrefmark{1}}
\IEEEauthorblockA{\IEEEauthorrefmark{1}Macquarie University, Sydney, Australia \\
Emails: \{osama.alhaddad, muhammad.ikram, young.lee\}@mq.edu.au}
\IEEEauthorblockA{\IEEEauthorrefmark{2}Data61, CSIRO, Sydney, Australia \\
Email: ejaz.ahmed@data61.csiro.au}
}

\begin{document}
\maketitle

\begin{abstract}

Security analysts are under growing pressure to assess increasingly complex and high-volume vulnerabilities. Recent advancements have shown that Large Language Models (LLMs) offer a promising solution by automating key aspects of this process through advanced natural language understanding. 
In this study, we evaluated four state-of-the-art LLMs—ChatGPT, Claude, Gemini, and DeepSeek—across twelve prompting techniques (PT) to assess their effectiveness in interpreting semi-structured and unstructured vulnerability information. 
As a practical use case, we investigate each model's ability to predict decision points within the Stakeholder-Specific Vulnerability Categorization (SSVC) framework, a widely used method for guiding response actions. Specifically, we focus on the four core SSVC decision points: Exploitation, Automatable, Technical Impact, and Mission \& Wellbeing. 

We leveraged a sample of 384 real-world vulnerabilities from the VulZoo dataset to evaluate the performance of each LLM across twelve PTs, including One-shot, Few-shot, and Chain-of-Thought. 
Over 165,000 LLM queries were issued to assess their predictive accuracy. Model responses were evaluated using F1-scores for each SSVC decision point (SDP), and weighted and unweighted Cohen’s Kappa metrics for SSVC decision outcomes (SDO). 
Our results show that Gemini consistently outperformed the other models, demonstrating superior performance on three out of four SDPs and yielding the highest number of correct recommendations. PTs that included exemplars generally led to improved accuracy, although all models exhibited difficulty with certain SDPs. Notably, only DeepSeek achieved fair agreement under weighted metrics, and all models showed a tendency to over-predict risk levels.

Our results suggest that while LLMs cannot yet replace expert judgment, certain LLM-PT combinations demonstrate moderate efficacy in specific decision areas in the SSVC vulnerability framework. When used appropriately, LLMs can serve as effective aids in vulnerability prioritization workflows, helping security analysts respond more efficiently to emerging threats.
\end{abstract}
\maketitle

\section{Introduction}
\label{sec:intro}
\input{files/intro}

\section{Background}
\label{sec:background}
\input{files/bg-rwork}

\section{Methodology}
\label{sec:estudy}
\input{files/estudy}

\section{Results and Analysis}
\label{sec:perevalresults}
This section presents the results obtained from our experiments and their relation to the research questions introduced in Section~\ref{sec:intro}. Comprehensive details and analyses of the experiments are available in Appendices~\ref{app:harmonic_means}, ~\ref{app:sdo_accuracy}, and~\ref{app:decision_tree_outcome}. The remainder of this section is structured to provide clear and detailed answers to each of the research questions outlined in Section~\ref{sec:intro}.

\subsection{\texorpdfstring{$\mathbf{RQ1}$}: Which LLM best analyzes vulnerability data to accurately recommend SDP responses?}
\label{subsec:rq1_discussion}

\begin{figure}[!th]
    \centering
    \includegraphics[width=0.9\columnwidth]{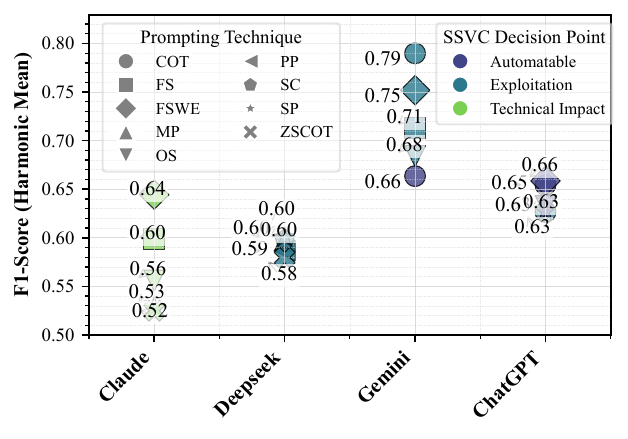}
    \caption{Top five performing PT and SDP combinations per LLM. See Appendix~\ref{app:harmonic_means} for the full dataset.}
    \label{fig:rq1_top_five_llm}
    \vspace{-0.45cm}
\end{figure}

To evaluate each LLM's effectiveness in vulnerability assessment, we identified the top five performing PT and SDP combinations based on harmonic mean F1-scores. As shown in Figure~\ref{fig:rq1_top_five_llm}, {Gemini consistently outperformed other models across all four SDPs}, with especially strong results in \textit{Exploitation} (0.79, CoT) and \textit{Automatable} (0.66, CoT). The model also achieved the highest scores for \textit{Technical Impact} (0.65, Self-refine (SR)) and \textit{Mission \& Wellbeing} (0.43, Memetic Proxy (MP)). These findings align with prior research that highlights Gemini's robust performance in text classification and summarization tasks~\cite{rai_msr78_2024, ferraguse}. Notably, the CoT prompting technique was highly effective for Gemini, reinforcing its ability to engage in structured, multi-step reasoning.

In comparison, ChatGPT demonstrated moderate but consistent performance across multiple SDPs. It excelled on \textit{Exploitation} and \textit{Automatable}, particularly with CoT and FS prompting. This aligns with earlier findings from Isogai et al.~\cite{isogai2024toward}, who reported ChatGPT’s moderate success in classifying CVSS sub-metrics, many of which map conceptually to SSVC's Exploitation and Automatable categories. Similar to Gemini, ChatGPT performed moderately with PTs such as OS and FSWE, but also with Self-planning (SP), suggesting it may require less complex reasoning scaffolds compared to Gemini or DeepSeek to reach effective conclusions.

Claude performed more modestly. Its strongest results hovered near 0.60 F1-score and were generally achieved through FS and CoT prompting. These results echo Krag et al.'s findings~\cite{krag2024large}, where Claude underperformed relative to its higher-end counterparts in document classification tasks. Although Claude was capable of producing coherent responses, it showed less consistency across different SDPs and PTs, suggesting limitations in its reasoning robustness under multi-context prompts.

DeepSeek R1, while recently released and showing theoretical parity with premium models, exhibited more variability in performance. It scored best on the \textit{Exploitation} SDP when paired with reasoning-based PTs such as SP and Self-consistency (SC), approximately 0.60. This aligns with DeepSeek's design, which incorporates explicit self-verification and error correction mechanisms~\cite{DeepSeek7:online}. However, its overall results trailed Gemini and ChatGPT. Notably, DeepSeek's highest scores were achieved with PTs that mirrored its internal architecture--particularly those that encouraged decomposition and iterative reasoning--suggesting that model-specific prompting alignment may be key to optimizing performance.

\textit{PT Insights.} CoT prompting emerged as the most effective technique across multiple models and SDPs, particularly for Gemini and ChatGPT. FSWE and SC also showed value when paired with LLMs capable of reflective reasoning (e.g., DeepSeek). In contrast, simpler PTs such as ZS or Priming tended to yield weaker results, reinforcing the importance of tailored reasoning scaffolds when eliciting structured outputs from LLMs. 

In a study investigating prompting LLMs to generate unsafe content, Arrieta et al. noted that less performant LLMs were less likely to generate unsafe content than DeepSeek \cite{arrieta2025o3}. Where DeepSeek performed strongest was on the Exploitation SDP when using the MP, SC, and SP PTs, scoring approximately 0.6. Mondillo et al.~\cite{mondillo2025comparative} observed that when comparing DeepSeek to ChatGPT's o1 on a multiple-choice question task in a medical setting, a Cohen's Kappa score of 0.2 indicated low agreement between the two models. The authors suggest that the approaches of the models could be an explanatory factor. ChatGPT o1 incorporates a CoT approach, while DeepSeek uses a self-reflection approach, which appears consistent with the results of this study. DeepSeek performed optimally on three of the four SDPs when using SC or SP, which may align better with its use of a \textit{\textless think\textgreater} tag to decompose problems and its self-verification mechanisms to identify and correct errors~\cite{DeepSeek7:online}.


\textbf{Takeaway:} {Gemini exhibited the most consistent and robust performance across all SDPs, particularly when paired with CoT and other exemplar-based PTs. While ChatGPT showed solid generalist capabilities, DeepSeek and Claude demonstrated narrower strengths, highly dependent on prompting alignment. These findings emphasize the importance of LLM-PT compatibility and suggest that careful pairing can significantly enhance vulnerability analysis outcomes.}

\subsection{\texorpdfstring{$\mathbf{RQ2}$}: Which LLM's SDPs, when parsed through the SSVC decision tree, most accurately recommend SDOs?}
\label{subsec:rq2_discussion}

To evaluate how effectively each LLM-PT pairing translates SDP predictions into actionable prioritization guidance, we parsed both LLM-generated and Vulnrichment-provided SDPs through the SSVC decision tree~\cite{noauthor_stakeholder-specific_nodate}. This yielded SSVC stakeholder decision outcomes (SDOs), which were then analyzed under a \textit{High} Mission \& Wellbeing Stand-in (MWS)--representative of high-stakes environments such as national infrastructure or safety-critical systems.

\begin{figure}[h!]
    \centering
    \includegraphics[width=\columnwidth]{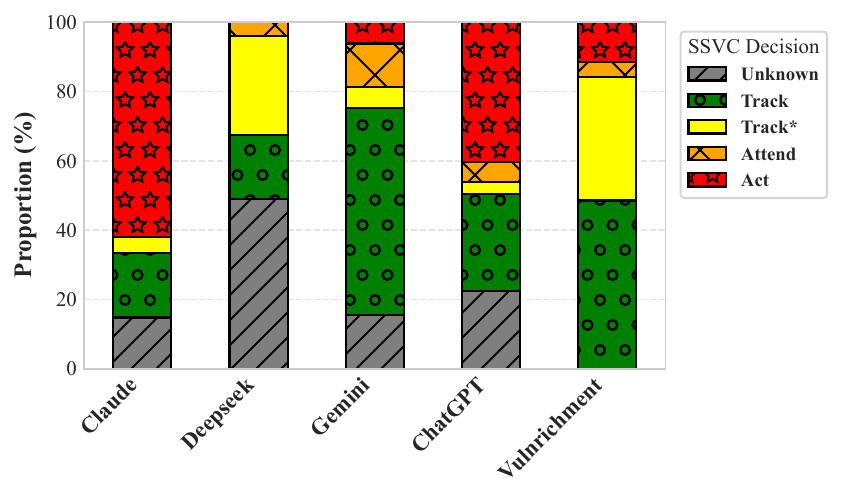}
    \caption{SDOs for the \textit{High} MWS. SDOs of less than 1\% omitted. See Appendix~\ref{app:decision_tree_outcome} for the full dataset.}    \label{fig:rq1_ssvc_high_mw_outcome}
    \vspace{-0.45cm}
\end{figure}

\textbf{SDO Distribution Alignment.}
Figure~\ref{fig:rq1_ssvc_high_mw_outcome} presents the proportional distribution of SDOs produced by each LLM compared to the Vulnrichment ground truth. Notably, Vulnrichment emphasizes SSVC's core design principle: reduce analyst burden by prioritizing only the most critical vulnerabilities for immediate remediation~\cite{SSVCASma50:online}. This is reflected in its outcome distribution—dominated by \textit{Track} (low priority) and tapering down through \textit{Track*} and \textit{Attend}, with a minimal proportion of \textit{Act} outcomes.

Among the LLMs, Gemini most closely approximates this prioritization profile. It generates the highest proportion of \textit{Track} outcomes and maintains comparable ratios of \textit{Act} and \textit{Attend} cases, suggesting an awareness of SSVC's intent to avoid over-prioritization. Conversely, other LLMs—namely {DeepSeek}, {ChatGPT}, and {Claude}--show inflated rates of high-urgency decisions (\textit{Act} and \textit{Attend}), at times exceeding the combined total of lower-urgency decisions (\textit{Track} and \textit{Track*}).

This over-prioritization behavior mirrors common pitfalls of traditional vulnerability prioritization frameworks like CVSS, where high severity is often over-attributed to vulnerabilities unlikely to be exploited~\cite{WhyYouNe59:online}. As a result, these LLMs risk replicating known shortcomings by recommending resource-intensive actions for vulnerabilities with limited practical risk.

\begin{figure}[!th]
    \centering
    \includegraphics[width=0.95\columnwidth]{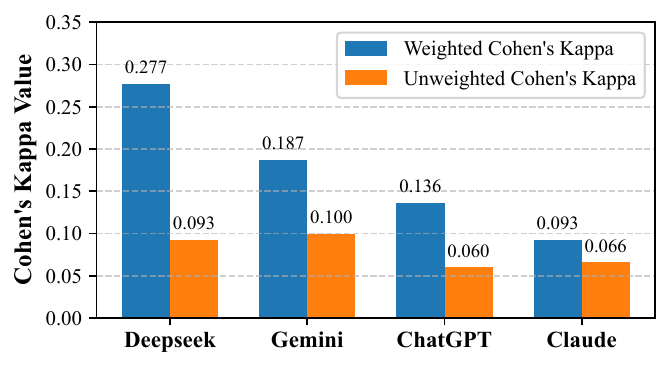}
    \caption{Unweighted and Weighted Cohen's Kappa scores for agreement between LLM SSVC decision outcomes and Vulnrichment ground truth SSVC decision outcomes for a \textit{High} Mission \& Wellbeing stand-ins (MWS).}
    \label{fig:ssvc_outcome_cohens_kappa}
    \vspace{-0.45cm}
\end{figure}

\textbf{Cohen's Kappa Agreement.}
We further evaluated outcome-level consistency using both Unweighted and Weighted Cohen's Kappa metrics, as illustrated in Figure~\ref{fig:ssvc_outcome_cohens_kappa}. Unweighted scores reflect exact match agreement, while Weighted scores account for the ordinal nature of SSVC outcomes, granting partial credit for near misses (e.g., predicting \textit{Attend} instead of \textit{Act}).

Across the board, {\it Unweighted} Cohen's Kappa scores were low, ranging from 0.06 to 0.10, indicating ``none'' to ``slight'' agreement~\cite{McHugh2012}. Gemini achieved the highest unweighted agreement, reinforcing its relative strength in matching the distributional profile of Vulnrichment.

In contrast, {DeepSeek (R1)} exhibited the highest {\it Weighted} Cohen's Kappa, reaching the ``fair'' agreement threshold. This suggests that although DeepSeek often misclassified SDOs by SSVC standards, its predictions were generally directionally correct—erring within one level of the correct outcome. This aligns with its architecture, which emphasizes internal self-verification and reflective problem decomposition~\cite{DeepSeek7:online}, and with its superior performance under SC and SP PTs.

Overall, the results show that accurate SDP classification alone is not sufficient to produce reliable SDOs. While Gemini (Flash 1.5) performed consistently well across both SDP and SDO evaluations, other models like DeepSeek showed isolated strengths--particularly when the error tolerance for ordinal misclassification was factored in. This suggests that prompting techniques aligned with an LLM's internal reasoning architecture (e.g., CoT for Gemini, self-reflection for DeepSeek) can improve prioritization fidelity, but only when used holistically across all four SDPs.

\textbf{Takeaway:} {Gemini yielded the most distributionally aligned SDO profile relative to Vulnrichment under a High MWS, and demonstrated the strongest unweighted agreement. DeepSeek, while less accurate in raw agreement, achieved fair performance under Weighted Cohen's Kappa, indicating relative alignment in prioritization direction. These findings underscore the need for comprehensive evaluation pipelines that go beyond raw accuracy and account for ordinal reasoning and practical prioritization utility.}

\subsection{\texorpdfstring{RQ3}: How does the choice of PT influence the performance of the LLMs? }
\label{subsec:rq3_discussion}

We investigated the impact of different PT on LLMs performance by comparing the F1-Scores achieved by each LLM across twelve PTs for the four SDPs~\cite{university_of_firat_institute_of_science_elazig_turkey_comparison_2022, noauthor_stakeholder-specific_nodate, TheSSVCm81:online}. The F1-score, a harmonic mean of precision and recall, was used to evaluate the performance of each LLM-PT-SDP combination across three independent trials~\cite{noauthor_we_nodate}.

\begin{figure}[!th]
    \centering
    \includegraphics[width=0.95\columnwidth]{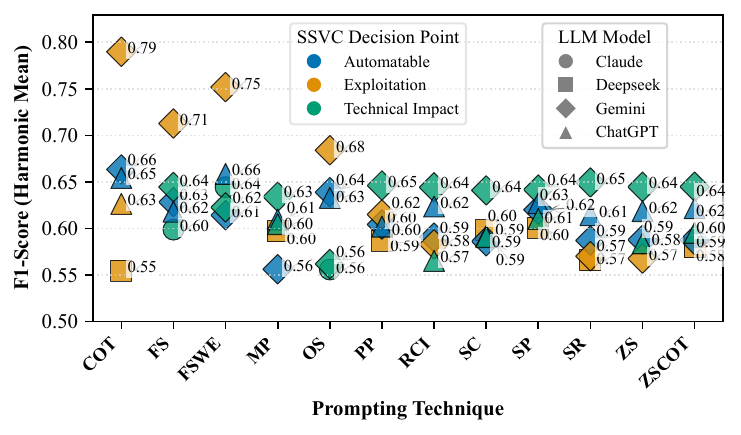}
    \caption{Top five performing LLM and SDP combinations per PT. Values rounded to 2 decimal places.} 
    \label{fig:rq2_top_five_pt}
\end{figure}

Figure \ref{fig:rq2_top_five_pt} presents a comparative overview of the top-performing LLM and SDP combinations for each PT.  Our analysis reveals a clear trend: PTs that incorporate exemplars generally outperform those that do not. CoT prompting achieved the highest overall F1-score, reaching 0.79 on the Exploitation SDP, demonstrating its effectiveness in facilitating complex reasoning.  FSWE followed closely with an F1-score of 0.75, also within the Exploitation SDP, underscoring the value of providing both examples and explanations to guide the LLM.

Interestingly, FS and OS prompting also exhibited strong performance, both peaking at the Exploitation SDP. This consistency across COT, FSWE, FS, and OS highlights a crucial factor: the presence of exemplars. These top four techniques all leverage in-context learning by providing the LLM with at least one demonstration, enabling it to better understand the task and desired output format. The Exploitation SDP, in particular, appears to benefit significantly from exemplar-based prompting, suggesting that this decision point may involve more complex reasoning that is effectively guided by examples.

The top four PTs share a key feature: they all include \textit{exemplars}--concrete input–output demonstrations embedded in the prompt. These serve to guide LLMs by reducing ambiguity, anchoring response structure, and offering implicit cues on how to reason through tasks. This aligns with best practices in prompt engineering, which emphasize that example-driven demonstrations help LLMs internalize task structure, especially for complex or nuanced outputs~\cite{bsharat_principled_2023}.

In contrast, PTs that do not include exemplars generally showed lower performance ceilings. For instance, SC achieved a maximum F1-score of 0.64. This aligns with findings by Gaur et al., who observed that while SC can enhance LLM reliability, it may also introduce spurious reasoning paths that require further refinement.  Similarly, SR and Zero-shot Chain-of-Thought (ZSCOT), despite their focus on reasoning, did not reach the performance levels of exemplar-based methods, likely due to the absence of concrete examples to ground the LLM's reasoning process. ZS prompting, which relies solely on instructions, also falls into this lower-performing category, further supporting the importance of exemplars. 

{\bf LLM–PT Synergy Matters.} While exemplar-based prompting generally performed best, the effectiveness of a given PT still depended on the model and the specific SDP. For instance, FSWE yielded strong results for both Gemini and ChatGPT on \textit{Automatable} and \textit{Exploitation}, but failed to match that performance on \textit{Mission \& Wellbeing}, where interpretive and contextual judgment is more complex. Similarly, the SR, SC and SP techniques yielded moderate performances for DeepSeek, whose internal architecture supports self-verification and iterative refinement~\cite{DeepSeek7:online}, even though it lacked exemplars from these prompts.

{\bf Complexity vs. Clarity Tradeoff.} Techniques such as SP, which introduces complex multi-step reasoning or problem decomposition, generally underperformed in our study. These PTs require the LLM to structure and verify its own logic, which may introduce variability without external cues. This reinforces a broader insight: \textit{prompt complexity alone is insufficient to guarantee performance}; rather, clarity and alignment with the model’s inductive biases—especially via exemplar scaffolding—are essential.

\textbf{Takeaway:} {PTs that incorporate exemplars--such as CoT, FSWE, FS, and OS—consistently yield stronger performance across LLMs and SDPs. Reasoning-based techniques without exemplars, while promising in theory, often underperform in practice unless carefully aligned with the model's architecture and task complexity.}


\subsection{\texorpdfstring{$\mathbf{RQ4}$}: Which SDP sees the strongest performance from an optimal combination of LLM and PT? }
\label{subsec:rq4_discussion}

\begin{figure}[!th]
    \centering
    \vspace{-0.15cm}
    \includegraphics[width=0.9\columnwidth]{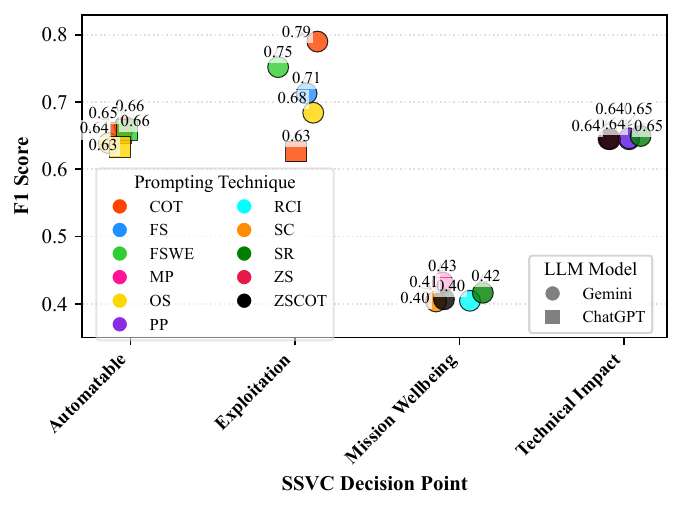}
    \vspace{-0.30cm}
    \caption{Top five LLM and PT combinations per SDP. Values rounded to 2 decimal places. }
    \label{fig:rq3_top_five_sdp}
    \vspace{-0.45cm}
\end{figure}

To answer this question, we identified the top five performing LLM-PT combinations by the harmonic mean of their F1-scores across three trials, grouped by each SSVC decision point (SDP). The results are summarized in Figure~\ref{fig:rq3_top_five_sdp}.

\textbf{Strongest Performing SDP: Exploitation.} Among all SDPs, the {Exploitation} category exhibited the strongest performance. The top combinations—predominantly involving {Gemini} paired with exemplar-driven techniques such as CoT, FSWE, and FS-achieved harmonic mean F1-scores ranging from {0.68 to 0.79}. This result is consistent with prior $\mathbf{RQ_1}$ and $\mathbf{RQ_2}$ analyses, where exemplar-based prompting yielded superior outcomes.
The superior performance on Exploitation can be attributed to both training data quality and the structured vulnerability information provided in the prompt. Exploitation-related attributes—such as presence in public exploit databases, descriptions in CVE/CVSS records, and indicators of compromise—are well-documented and widely available across multiple cybersecurity knowledge bases~\cite{bullough2017predicting}. LLMs likely encountered similar representations during pretraining, enabling more confident inference from the input prompts~\cite{Singhal2023}.

\paragraph{Moderately Performing SDPs: Automatable and Technical Impact.}
The next best performing SDPs were {Automatable} and {Technical Impact}, with maximum F1-scores in the range of {0.63 to 0.66}. These were achieved by combinations involving 4o-mini and Flash 1.5 under FS, FSWE, and SC PTs.

These SDPs rely on the LLM's ability to infer operational feasibility (e.g., whether an attack can be scripted) and technical consequence (e.g., effect on confidentiality, integrity, availability). Although these properties are present in structured sources like CVSS vectors, they often require deeper contextual interpretation—e.g., recognizing from a technical description that a buffer overflow enables remote code execution. While LLMs performed reasonably well, performance was slightly limited by variability in how these traits are described in free-text fields of vulnerability datasets.

\textbf{Lowest Performing SDP: Mission \& Wellbeing.}
 The Mission \& Wellbeing SDP demonstrated consistently lower performance across all LLMs and PTs. Specifically, the highest observed F1-scores were constrained to a range between 0.40 and 0.43, with the Gemini model generally achieving these peak values. Several contributing factors can explain this relatively weaker performance. First, the Vulnrichment dataset lacks inherent ground-truth values for Mission \& Wellbeing. To address this, the research employed synthetic Mission \& Wellbeing Stand-ins (MWSs), designed to represent varying levels of organizational risk (high, medium, and low). While these stand-ins aimed to offer contextual cues, they may have lacked the necessary granularity for precise interpretation by the LLMs. Second, the very nature of the Mission \& Wellbeing SDP introduces inherent contextual complexity. Unlike other SDPs, evaluating Mission \& Wellbeing necessitates an understanding of an organization's mission, a factor that typically hinges on nuanced human knowledge regarding infrastructure, stakeholder dynamics, and broader societal implications~\cite{Limitati10:online}. Consequently, the absence of rich contextual information likely compelled LLMs to rely on more heuristic-driven estimations. Third, the design of the prompts themselves might have introduced bias. The prompts used throughout this study maintained a fixed order for presenting options, a factor that could have influenced the selection process. Prior research has indeed indicated that LLMs can exhibit position bias, demonstrating a tendency to favor options appearing either at the beginning or the end of ordered lists~\cite{pezeshkpour_large_2023, lican2024}.

These findings highlight the challenges of encoding high-level organizational context into prompts that remain usable by LLMs. Developing richer, domain-specific context scaffolds for Mission \& Wellbeing may be critical for future research and operational deployment.

\textbf{LLM–PT Sensitivity.} Across all SDPs, the results reaffirm the importance of aligning PT with model architectures. Gemini performed best when paired with reasoning- or exemplar-rich techniques, while DeepSeek benefited from iterative refinement techniques such as SC and SR. However, no model demonstrated strong performance across all SDPs, suggesting that adaptive prompting--where techniques are dynamically selected based on the SDP--may yield better results in practical deployments.

\textbf{Takeaway:} {The Exploitation SDP exhibited the highest performance under optimal LLM–PT combinations, likely due to the abundance of high-quality training and input data. In contrast, Mission \& Wellbeing remained the most challenging, due to the need for deep contextual understanding and limitations in synthetic risk framing. These results underscore the need for SDP-specific prompting techniques and richer context integration for improved prioritization accuracy.}

\subsection{\texorpdfstring{$\mathbf{RQ5}$}: Do LLMs suffer from recommending false positive or false negative SDOs?}
\label{subsec:rq5_discussion}

To address this question, we first established clear definitions for false positives and false negatives within the context of our study. Specifically, we defined false positives as instances where the LLM recommends a higher-severity outcome than the ground truth, while false negatives are cases where the LLM suggests a lower-severity outcome. To illustrate, consider a scenario where the Vulnrichment SDO for a given vulnerability is \textit{Act}, but the LLM SDO recommends \textit{Track}. This constitutes a false negative, potentially leading a security analyst to underestimate the risk and forgo necessary remediation. Conversely, if the Vulnrichment SDO is \textit{Track} and the LLM SDO recommends \textit{Attend}, this false positive could result in the inefficient allocation of resources to address a vulnerability that may not pose an immediate threat. Following these definitions, we proceeded to compare the SSVC SDOs generated by parsing the LLM-predicted SDP responses through the SSVC decision tree against the \textit{ground truth} SDOs derived from Vulnrichment.

These error types have different operational implications. False positives may lead to the unnecessary allocation of resources toward low-risk vulnerabilities, while false negatives risk deprioritizing vulnerabilities that require urgent attention, potentially exposing systems to unmitigated threats.

\begin{figure}[th]
    \centering    
    \vspace{-0.15cm}
    \includegraphics[width=0.95\columnwidth]
    {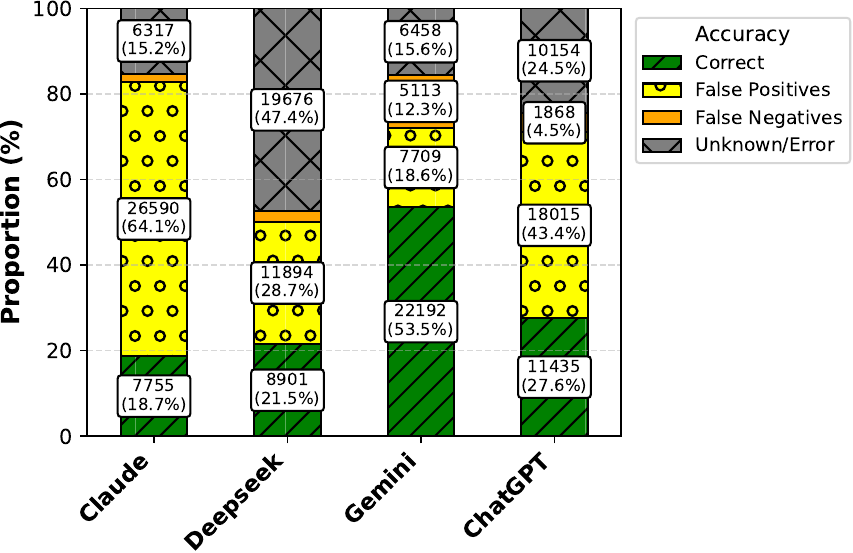}
    \vspace{-0.25cm}
    \caption{Accuracy bar chart for LLM SDOs. When LLM SDP responses are parsed through the SSVC decision tree, the LLM SDOs are compared against the Vulnrichment ground truth SDOs. If an LLM SDO is higher in severity than the ground truth (e.g., LLM SDO is \textit{Act} while ground truth SDO is actually \textit{Track}), this is classified as a false positive. Conversely, an LLM SDO of lower severity than the ground truth SDO is a false negative. }
    \label{fig:rq4_llm_accuracy}
    \vspace{-0.25cm}
\end{figure}

\textbf{Error Distribution Across Models.} As illustrated in Figure~\ref{fig:rq4_llm_accuracy}, all LLMs tended to generate false positive SDOs. For each model, the number of false positives exceeded the false negatives. This over-prediction trend mirrors findings in prior literature on LLM behavior in security-sensitive tasks. For example, Basic et al.~\cite{basic2024large} observed that LLMs in vulnerability detection tasks exhibit high false positive rates due to their sensitivity to incomplete contextual cues. Steenhoek et al.~\cite{steenhoek2024comprehensive} reported that LLMs often misinterpret missing null or boundary checks as vulnerabilities, leading to over-reporting. Similarly, Purba et al.~\cite{10301302} noted that LLMs scored higher false positive rates than proprietary scanners in SQL injection detection, frequently failing to recognize input sanitization.

Among the models studied, Gemini was the only LLM to generate more correct classifications than false positives and false negatives, indicating a relatively stronger ability to identify vulnerabilities accurately. However, it also exhibited a disproportionately higher false negative rate compared to other models. This behavior is consistent with findings by Masood et al.~\cite{masood2024beyond}, who reported that Gemini tended to miss subtle vulnerabilities in complex technical tasks. The model’s conservative classification thresholds may contribute to both fewer over-predictions and a higher likelihood of under-prioritization.

\textbf{Implications of False Positive Dominance.} The prevalence of false positives across models suggests that LLMs are inclined to {overestimate vulnerability severity}, potentially due to risk-averse generation patterns or the influence of training data skewed toward high-impact vulnerabilities. While such conservatism may be desirable in certain security domains (e.g., malware detection), it imposes a cost in vulnerability management, where analyst time and remediation effort are limited resources. Over-prioritizing low-impact vulnerabilities can divert attention from genuinely critical ones.

Conversely, false negatives pose a more severe risk but occur less frequently. Nonetheless, the elevated FN counts from Gemini--despite its otherwise strong performance--indicate that even high-performing models may suffer from under-detection biases, particularly when task complexity increases or when contextual clarity is low.

\textbf{Model-Specific Error Profiles.}
It is important to note that the types and magnitudes of errors varied across the LLMs. For instance, Gemini exhibited the highest count of correct classifications, however, it also recorded the highest false negative count, suggesting a profile characterized by strong precision but only moderate recall. In contrast, ChatGPT, Claude, and DeepSeek showed a tendency for false positives to significantly outnumber correct classifications, indicating potentially poor calibration and a high degree of overestimation. This observed variation across models underscores the importance of considering not only average performance metrics when selecting LLMs, but also the operational risks associated with their specific and distinct error profiles.

\textbf{Takeaway:} {All LLMs in this study demonstrated a consistent tendency to over-prioritize vulnerabilities, resulting in high false positive rates. Gemini, while yielding the highest number of true positives, also produced more false negatives than other models, suggesting a trade-off between precision and recall. These findings highlight the need for calibration mechanisms and error-aware model selection in vulnerability triage workflows.}

\section{Assumptions and Limitations}

\textbf{Data Leakage.}
While this study did not involve the use of confidential or proprietary organizational data, it is important to acknowledge the privacy and security implications in real-world deployments. In practice, security analysts may be hesitant to submit sensitive infrastructure or business-impact information to proprietary LLM APIs hosted by third-party vendors. This introduces the risk of \textit{inference leakage}, where the LLM provider could potentially retain, analyze, or inadvertently reveal confidential content to other users through prompt injection or model inversion attacks~\cite{LLMInfor79:online}. Mitigating this risk requires the adoption of robust organizational {data governance frameworks}, specifying what types of data are permissible for third-party LLM use and enforcing appropriate redaction, anonymization, or deployment on private infrastructure~\cite{Datagove49:online}.

\textbf{Data Integrity.}
The reliability of machine learning models is inherently dependent on the integrity of their training and input data. A critical but underexplored threat is \textit{data poisoning}, where adversaries introduce corrupted or misleading content into public vulnerability datasets. Such poisoning could mislead both human analysts and automated systems into under- or overestimating the criticality of specific vulnerabilities~\cite{Das2024.03.20.24304627}. While we assumed the VulZoo and Vulnrichment datasets used in this study were free from tampering, this cannot be definitively guaranteed. Future applications of LLMs in vulnerability analysis should incorporate anomaly detection and source verification mechanisms to guard against such risks.

\textbf{Data Bias.}
A further limitation arises from potential bias in the evaluation dataset. Vulnrichment contains relatively recent vulnerability disclosures, with a median age of three months~\cite{noauthor_we_nodate}, whereas the broader CVE corpus from MITRE spans vulnerabilities with a median age closer to three years. This recency bias may favor LLMs trained on recent corpora and limit generalizability to legacy systems or older software stacks. As such, an important area for future work involves evaluating LLMs on temporally diverse datasets to better understand performance consistency over time.

\section{Conclusion and Future Work}

This study presented a comprehensive empirical evaluation of four LLMs across twelve PTs, to assess their effectiveness in recommending SDPs and SDOs for real-world software vulnerabilities. Our key findings indicate that Gemini demonstrated the strongest overall performance, achieving the highest F1-scores across all four SDPs and generating SDO distributions most closely aligned with ground truth. Furthermore, exemplar-based PTs, including CoT, FS, and FSWE, consistently yielded superior results, emphasizing the importance of contextual demonstrations in guiding LLM reasoning. The Exploitation and Automatable SDPs proved more tractable for LLMs, likely due to richer training and prompt data; conversely, the Mission \& Wellbeing SDP presented a greater challenge due to its reliance on nuanced organizational context that was difficult to capture using generic stand-in scenarios. Across all LLMs, false positive predictions significantly outnumbered false negatives, revealing a bias toward over-prioritization, and while Gemini showed the highest number of correct classifications, it also exhibited more false negatives, illustrating a trade-off between precision and recall. Ultimately, LLMs still face challenges in SDO generation when relying solely on parsed SDP inputs, particularly for context-sensitive decisions, suggesting that while they can assist analysts, they are not yet capable of fully replacing expert judgment in high-stakes vulnerability triage.

Our findings highlight both the potential and the limitations of LLMs within security analyst workflows. While the appropriate combination of LLM and PT can achieve moderate performance for specific SDPs, the complete automation of vulnerability prioritization remains an unrealized goal. Future work aimed at improving performance on SDO prediction could explore several avenues: optimizing multiple-choice prompt design, including randomized option ordering, to mitigate positional bias; capturing and encoding detailed organizational risk profiles to enable more accurate Mission \& Wellbeing assessments; and investigating fine-tuning or RAG (Retrieval-Augmented Generation) based approaches to infuse LLMs with trusted, up-to-date context while adhering to data privacy regulations.

\balance

\bibliographystyle{IEEEtran}
\bibliography{references}

\appendices
\onecolumn  

\section{Stakeholder-Specific Vulnerability Categorization Decision Points (SDP)}
\label{app:sdp}
\begin{table}[H]  
    \small  
    \centering
    \begin{tabular}{p{3cm}|p{4cm}|p{9.5cm}}
        \toprule
        \hline
        \textbf{Decision Point} & \textbf{Description} & \textbf{Options} \\
        \hline
        Exploitation & Evidence of active exploitation of a vulnerability. & \textbf{None:} No evidence of active exploitation and no public proof of concept (PoC). \newline \textbf{Public PoC:} Either (1) a public PoC exists in sources like Metasploit, or (2) the vulnerability has a well-known method of exploitation. \newline \textbf{Active:} Credible public evidence that threat actors have used the exploit in the wild. \\
        \hline
        Technical Impact & Systems impacts of exploiting a vulnerability. & \textbf{Partial:} Limited control or exposure of the software behavior, low chance of total control. \newline \textbf{Total:} The exploit gives total control or total disclosure of information on the system. \\
        \hline
        Automatable & Ease and speed of automating exploitation events. & \textbf{No:} Steps 1-4 of the kill chain cannot be reliably automated. \newline \textbf{Yes:} Steps 1-4 of the kill chain can be reliably automated, such as for unauthenticated RCE or command injection vulnerabilities. \\
        \hline
        Mission \& Wellbeing & Impact on organization functions, systems, and human health and safety. & \textbf{High:} Irreversible impact on corporate functions OR irreversible impact on human health and safety. \newline \textbf{Medium:} Moderate impact on corporate support functions AND minimal or moderate impact on human health and safety. \newline 
        \textbf{Low:} Minimal impact on corporate functions AND minimal impact on human health and safety. \\
        \hline
        \bottomrule
    \end{tabular}
    \caption{SSVC decision points for vulnerability prioritization. Taken and adapted from the CISA stakeholder-specific vulnerability categorization Guide~\cite{noauthor_cisa_nodate} (cf. \S\ref{subsec:ssvc}).}
    \label{tab:ssvc-decision-points}
\end{table}

\section{Stakeholder-Specific Vulnerability Categorization Decision Outcomes (SDO)}
\label{app:sdo}
\begin{table}[H]  
    \small  
    \centering
    \begin{tabular}{p{5cm}|p{11.5cm}}
        \toprule
        \hline
        \textbf{Outcome} & \textbf{Description} \\
        \hline
        Track & No immediate action required. Monitor and reassess if new information emerges. Remediate within standard update timelines. \\
        \hline
        Track* & No immediate action is required. Monitor closely due to specific characteristics. Remediate within standard update timelines. \\
        \hline
        Attend & Involve supervisors. Initiate assistance requests and notifications if necessary. Remediate sooner than standard timelines. \\
        \hline
        Act & Involve supervisors and leadership. Requires immediate response planning and execution. Remediate as soon as possible. \\
        \hline
        \bottomrule
    \end{tabular}
    \caption{SSVC decision outcomes (SDOs) taken and adapted from the CISA stakeholder-specific vulnerability categorization guide~\cite{noauthor_cisa_nodate} (cf. \S\ref{subsec:ssvc}).}
    \label{tab:ssvc_outcomes}
\end{table}

\section{Mission \& Wellbeing Stand-ins (MWS)}
\label{app:mission_wellbeing_standins}
\begin{table}[H]  
    \small  
    \centering
    \begin{tabular}{p{3cm}|p{4cm}|p{9.5cm}}
        \toprule
        \hline
        \textbf{Risk Scenario} & \textbf{Risk Description} & \textbf{Reasoning} \\
        \hline
        High & \textit{Assume you work for an electric power generation company that is the sole provider for the country.} & Power generation for a country is critical to maintaining national economic and social interests. Loss of power can disrupt key economic and social services and can have a severe impact on health and safety. \\
        \hline
        Medium & \textit{Assume you work for a supermarket chain that operates outlets across the state.} & Inability to provide key supplies to society in an accessible manner, including food and basic medicine, can impact health and safety. \\
        \hline
        Low & \textit{Assume you work for a small town leisure centre that provides sport and recreational facilities and services to the community.} & While an inconvenience to a community, a closure of a leisure centre is unlikely to pose a risk to health and safety. \\
        \hline
        \bottomrule
    \end{tabular}
    \caption{Mission \& Wellbeing stand-in scenarios to gauge the ability of LLMs to be aware of the organization's context and the impact of vulnerabilities (cf. \S\ref{subsubsec:missing_mw}, \S\ref{subsec:rq4_discussion}).}
    \label{tab:mission_wellbeing_standins}
\end{table}

\section{Prompting Categories \& Techniques}
\label{app:prompt_categories_techniques}
\begin{table}[ht]
\label{tab:prompting_techniques}
\begin{adjustbox}{max width=0.95\textwidth, max height=1\textheight}
\begin{tabular}{p{0.2\textwidth}p{0.9\textwidth}}
\toprule
\textbf{Prompting Category \newline \& Technique} & \textbf{Description} \\
\midrule
\textbf{Root} & Instructs an LLM to complete a task accompanied by varying numbers of examples, ranging from providing no prior example (i.e., Zero-shot), providing one prior example (i.e., One-shot), and providing several prior examples (i.e., Few Shot). \\
\addlinespace
Zero-shot (ZS) & The prompt does not provide the LLM with any supporting information or worked example, forcing the LLM to primarily rely on its prior training to generate the desired outcome. LLM performance tends to be poorer than Few-shot prompting, as ``without few-shot exemplars, it is harder for models to perform well on prompts that are not similar to the format of the pretraining data"~\cite{wei2021finetuned}. While seemingly sub-optimal, given a vulnerability may be new with limited data, a Zero Shot prompt can still be an appropriate prompting technique to generate a recommended response. \\
\addlinespace
One-shot (OS) and Few-shot (FS) & These prompting techniques include examples as part of the instruction; One-shot provides one example and Few-shot several~\cite{noauthor_prompt_nodate}. Research notes that ``the main advantage of few-shot is a major reduction in the need for task-specific data", though ``a small amount of task-specific data is still required"~\cite{brown_language_2020}. Given that the JSON files in the Vulnrichment dataset have varying degrees of details, providing one or several exemplars could be an appropriate technique to guide the LLMs where information is available. \\
\addlinespace
\textbf{Refinement-based} & Instructs an LLM to iterate, review, and provide feedback on its responses before returning an optimized response to the user. \\
\addlinespace
Recursive Criticism and Improvement (RCI) & This prompting approach comprises three sub-prompts: (1) the original prompt, where the user asks the LLM to complete a query; (2) the critique prompt, where the user asks the LLM to critically analyse the output; and (3) the improvement prompt, where the user asks the LLM to improve the response based on the critical analysis~\cite{kim_language_2023}. This prompting technique can assist the LLMs to critique their initial response and, if sub-optimal, regenerate a more optimal response. \\
\addlinespace
Self-refine (SR) & This prompt includes instructions to the LLM to generate a draft response, draft feedback on that response, and provide an improved final response, either in a nominated number of iterations or when the LLM is satisfied with its response~\cite{madaan_self-refine_2023}. As with RCI, this introspective feature could yield improved performance. However, given a time-pressured vulnerability prioritization environment, allowing an LLM to continue to iterate until it is satisfied with a response may result in excessively delayed responses. As such, a limit on iterations is required. \\
\addlinespace
\textbf{Decomposition-based} & Provides an LLM a breakdown of complex tasks into manageable subtasks, and instructs the LLM to process those subtasks before reconstituting those completed subtasks to return an optimized response. \\
\addlinespace
Self-planning (SP) & This prompt comprises two parts: (1) requesting the LLM to generate a plan to complete a given task; and (2) using the generated plan to complete the task~\cite{jiang_self-planning_2024}. While similar to Least-to-Most, minimal user input is required, as the LLM is breaking down the problem into sub-problems to solve, not the user. \\
\addlinespace
\textbf{Reasoning-based} & Instructs an LLM to generate lines of reasoning, explain those lines of reasoning, and follow those lines of reasoning to develop a response. \\
\addlinespace
Chain-of-Thought (CoT) & This prompt ``incorporate[s] intermediate steps to guide [a] progressive reasoning", typically comprising ``demonstrations and textual instructions"~\cite{yu_towards_2023}. This technique has been shown to yield improved ``performance on a range of arithmetic, commonsense, and symbolic reasoning tasks"~\cite{wei_chain--thought_2022}. For this prompting technique to work in this study, a mapping demonstration would be required, indicating to the LLM how, for example, the properties of a given vulnerability map to respective SDP values. \\
\addlinespace
Zero-shot Chain-of-Thought (ZSCOT) & This prompt encourages the LLM to generate an implicit reasoning process~\cite{kojima_large_nodate}. One study noted that by including the simple instruction ``let's think step by step" in their prompts, the authors were able to improve math reasoning abilities of GPT-3 from 17.7\% to 78.7\%~\cite{chen_when_2023}. \\
\addlinespace
Self-consistency (SC) & This prompt instructs the LLM to generate multiple responses for the task; using the Model Picks variant, the LLM determines which response was generated the most often and returns this to the user~\cite{ahmed_better_2023}. Assuming a Model Picks variant and not a User Picks, where the user must determine which of the responses to accept, this could help security analysts working to tight deadlines. \\
\addlinespace
Few-shot with Explanation (FSWE) & This prompt provides the LLM with input and output exemplars, followed by an explanation of the reasoning; the LLM uses this template to generate its answer, followed by an explanation of how it generated this answer~\cite{lampinen_can_2022}. As with CoT, several mapping exemplars can be provided, with the addition of demonstrating how the mapping is conducted. No further prompting iterations are required, and research indicates improved performance over Zero Shot and Few Shot~\cite{lampinen_can_2022}. \\
\addlinespace
\textbf{Priming} & Provides an LLM a context in which to process a subsequently provided task. \\
\addlinespace
Persona Pattern (PP) & This prompt instructs the LLM to adopt the viewpoint of a given person or role before completing a task, though this approach has not been evaluated in experiments~\cite{tony_prompting_2024}. For a security engineering environment, priming the LLM to position itself in this environment may be beneficial to improve vulnerability prioritization performances. \\
\addlinespace
Memetic Proxy (MP) & This prompt instructs the LLM to assume a given scenario or situation, as opposed to adopting a particular role~\cite{tony_prompting_2024}. As with Persona Pattern, this priming feature could improve the LLM's ability to prioritize vulnerabilities. \\
\bottomrule
\end{tabular}
\end{adjustbox}
\caption{PTs and their categories used in this study. Adapted from Tony et al. (cf. \S\ref{subsec:parse_sample}).}
\end{table}

\clearpage  
\section{Harmonic Means of trial F1-Scores}
\label{app:harmonic_means}
\begin{table}[ht]
\centering
\begin{adjustbox}{max width=0.82\textwidth, max height=0.7\textheight}
\begin{tabular}{ll|r|r|r|r}
\toprule
\textbf{\small Prompting Technique} & \textbf{\small SSVC Decision Point} & \multicolumn{1}{c|}{\textbf{\small Claude 3 Haiku}} & \multicolumn{1}{c|}{\textbf{\small Gemini Flash 1.5}} & \multicolumn{1}{c|}{\textbf{\small ChatGPT 4o-mini}} & \multicolumn{1}{c}{\textbf{\small DeepSeek-R1}} \\
& & \textbf{\footnotesize F1} & \textbf{\footnotesize F1} & \textbf{\footnotesize F1} & \textbf{\footnotesize F1} \\
\midrule
Chain-of-Thought & Automatable & 0.3062 & 0.6635 & 0.6543 & 0.3471 \\
& Exploitation & 0.0406 & 0.7898 & 0.6265 & 0.5545 \\
& Mission \& Wellbeing & 0.2267 & 0.3295 & 0.2523 & 0.2427 \\
& Technical Impact & 0.5224 & 0.5274 & 0.4606 & 0.4863 \\
\midrule
Few-shot & Automatable & 0.2794 & 0.6284 & 0.6181 & 0.3615 \\
& Exploitation & 0.0711 & 0.7128 & 0.4560 & 0.5109 \\
& Mission \& Wellbeing & 0.2511 & 0.3244 & 0.2388 & 0.2375 \\
& Technical Impact & 0.5988 & 0.6446 & 0.5220 & 0.4904 \\
\midrule
Few-shot with Explanation & Automatable & 0.3419 & 0.6144 & 0.6586 & 0.3889 \\
& Exploitation & 0.0927 & 0.7518 & 0.5854 & 0.5533 \\
& Mission \& Wellbeing & 0.2644 & 0.3168 & 0.2704 & 0.2488 \\
& Technical Impact & 0.6443 & 0.6227 & 0.5181 & 0.5043 \\
\midrule
Memetic Proxy & Automatable & 0.1416 & 0.5562 & 0.6096 & 0.2948 \\
& Exploitation & 0.0549 & 0.5446 & 0.1500 & 0.5970 \\
& Mission \& Wellbeing & 0.2321 & 0.4317 & 0.2503 & 0.2472 \\
& Technical Impact & 0.4889 & 0.6341 & 0.6048 & 0.5006 \\
\midrule
One-shot & Automatable & 0.2662 & 0.6390 & 0.6329 & 0.3133 \\
& Exploitation & 0.0289 & 0.6841 & 0.5017 & 0.5287 \\
& Mission \& Wellbeing & 0.2369 & 0.3261 & 0.2455 & 0.2382 \\
& Technical Impact & 0.5560 & 0.5619 & 0.5138 & 0.4789 \\
\midrule
Persona Pattern & Automatable & 0.1134 & 0.6043 & 0.6021 & 0.2908 \\
& Exploitation & 0.0429 & 0.6150 & 0.1664 & 0.5869 \\
& Mission \& Wellbeing & 0.2321 & 0.3937 & 0.2386 & 0.2428 \\
& Technical Impact & 0.4814 & 0.6461 & 0.5818 & 0.5028 \\
\midrule
Recursive Criticism \& Improvement & Automatable & 0.1298 & 0.5902 & 0.6236 & 0.2692 \\
& Exploitation & 0.0335 & 0.5830 & 0.1573 & 0.5229 \\
& Mission \& Wellbeing & 0.2434 & 0.4047 & 0.2327 & 0.2369 \\
& Technical Impact & 0.4687 & 0.6442 & 0.5653 & 0.4920 \\
\midrule
Self-consistency & Automatable & 0.1049 & 0.5862 & 0.5919 & 0.3116 \\
& Exploitation & 0.0165 & 0.5752 & 0.1538 & 0.5983 \\
& Mission \& Wellbeing & 0.2370 & 0.4037 & 0.2338 & 0.2563 \\
& Technical Impact & 0.4388 & 0.6409 & 0.5907 & 0.5355 \\
\midrule
Self-planning & Automatable & 0.1594 & 0.6201 & 0.6271 & 0.2906 \\
& Exploitation & 0.1043 & 0.5983 & 0.2640 & 0.6007 \\
& Mission \& Wellbeing & 0.2470 & 0.3908 & 0.2520 & 0.2459 \\
& Technical Impact & 0.5288 & 0.6418 & 0.6102 & 0.5102 \\
\midrule
Self-refine & Automatable & 0.1436 & 0.5871 & 0.6144 & 0.2747 \\
& Exploitation & 0.0597 & 0.5702 & 0.1321 & 0.5661 \\
& Mission \& Wellbeing & 0.2559 & 0.4165 & 0.2319 & 0.2365 \\
& Technical Impact & 0.5094 & 0.6495 & 0.5593 & 0.4845 \\
\midrule
Zero-shot & Automatable & 0.1390 & 0.5879 & 0.6187 & 0.2741 \\
& Exploitation & 0.0700 & 0.5677 & 0.1633 & 0.5627 \\
& Mission \& Wellbeing & 0.2356 & 0.4000 & 0.2338 & 0.2313 \\
& Technical Impact & 0.5207 & 0.6448 & 0.5841 & 0.4913 \\
\midrule
Zero-shot Chain-of-Thought & Automatable & 0.1496 & 0.5876 & 0.6218 & 0.2786 \\
& Exploitation & 0.0758 & 0.5789 & 0.1663 & 0.5797 \\
& Mission \& Wellbeing & 0.2504 & 0.4068 & 0.2385 & 0.2378 \\
& Technical Impact & 0.5249 & 0.6449 & 0.5954 & 0.5071 \\
\bottomrule
\end{tabular}
\end{adjustbox}
\caption{Harmonic means of the F1-scores from the three trials for LLM, PT and SDP (cf. \S\ref{subsec:rq1_discussion}, \S\ref{subsec:rq3_discussion}).}
\label{tab:llm_pt_sdp_performance}
\end{table}

\clearpage

\section{LLM SSVC Decision Outcome Accuracy}
\label{app:sdo_accuracy}
\begin{table}[H]
\centering
\label{tab:llm_performance}
\renewcommand{\arraystretch}{1.2} 
\setlength{\tabcolsep}{4pt} 
\begin{tabular}{l @{\hspace{1.5cm}} r @{\hspace{0.5cm}} r @{\hspace{0.5cm}} r @{\hspace{0.5cm}} r @{\hspace{0.5cm}} r}
\toprule
\textbf{LLM} & \textbf{Correct} & \textbf{False Negative} & \textbf{False Positive} & \textbf{Unknown/Error} & \textbf{Total} \\
\midrule
Claude 3 Haiku & 7,755 (15.42\%) & 810 (9.21\%) & 26,590 (41.36\%) & 6,317 (14.86\%) & 41,472 (100\%) \\
DeepSeek-R1 & 8,901 (17.70\%) & 1,001 (11.38\%) & 11,894 (18.50\%) & 19,676 (46.28\%) & 41,472 (100\%) \\
Gemini Flash 1.5 & 22,192 (44.14\%) & 5,113 (58.13\%) & 7,709 (11.99\%) & 6,458 (15.19\%) & 41,472 (100\%) \\
ChatGPT 4o-mini & 11,435 (22.74\%) & 1,868 (21.24\%) & 18,015 (28.03\%) & 10,154 (23.88\%) & 41,472 (100\%) \\
\bottomrule
\end{tabular}
\caption{LLM performance in accurately recommending SDOs. Percentages in parentheses represent the proportion of each classification type across all LLMs. A False Negative is where the LLM has recommended an SDO that is less severe than the Vulnrichment ground truth. A False Positive is where the LLM has recommended an SDO that is more severe than the Vulnrichment ground truth (cf. \S\ref{subsec:rq5_discussion}).}
\end{table}

\section{SSVC Decision Tree Outcome Analysis}
\label{app:decision_tree_outcome}
\begin{table}[H]
    \centering \label{tab:llm_sdp_responses_distributions}
    \begin{tabular}{|l|l|r|r|r|r|r|r|r|r|}
        \hline
        \multirow{2}{*}{\textbf{Model}} & \multirow{2}{*}{\textbf{SSVC Decision}} & \multicolumn{2}{c|}{\textbf{Low M\&W}} & \multicolumn{2}{c|}{\textbf{Medium M\&W}} & \multicolumn{2}{c|}{\textbf{High M\&W}} & \multicolumn{2}{c|}{\textbf{Subtotal}} \\
        \cline{3-10}
         &  & \textbf{Count} & \textbf{\%} & \textbf{Count} & \textbf{\%} & \textbf{Count} & \textbf{\%} & \textbf{Count} & \textbf{\%} \\
        \hline
        \multirow{6}{*}{Vulnrichment} 
            & Act     & 0     & 0.0\%   & 288   & 2.1\%   & 576    & 4.2\%   & 864    & 2.1\%   \\
            & Attend  & 432   & 3.1\%   & 288   & 2.1\%   & 1,584  & 11.5\%  & 2,304  & 5.6\%   \\
            & Track*  & 0     & 0.0\%   & 468   & 3.4\%   & 4,896  & 35.4\%  & 5,364  & 12.9\%  \\
            & Track   & 13,284& 96.1\%  & 12,672& 91.7\%  & 6,660  & 48.2\%  & 32,616 & 78.7\%  \\
            & Unknown & 108   & 0.8\%   & 108   & 0.8\%   & 108    & 0.8\%   & 324    & 0.8\%   \\
            & Error   & 0     & 0.0\%   & 0     & 0.0\%   & 0      & 0.0\%   & 0      & 0.0\%   \\
        \cline{2-10}
        \multicolumn{2}{|l|}{\textbf{Subtotal}} & 13,824 & 100.0\% & 13,824 & 100.0\% & 13,824 & 100.0\% & 41,472 & 100.0\% \\
        \hline
        \multirow{6}{*}{Claude 3 Haiku} 
            & Act     & 8,628  & 62.4\%  & 8,545  & 61.8\%  & 8,532  & 61.8\%  & 25,705 & 62.1\%  \\
            & Attend  & 484    & 3.5\%   & 534    & 3.9\%   & 642    & 4.7\%   & 1,660  & 4.0\%   \\
            & Track*  & 250    & 1.8\%   & 181    & 1.3\%   & 33     & 0.2\%   & 464    & 1.1\%   \\
            & Track   & 2,432  & 17.6\%  & 2,639  & 19.1\%  & 2,579  & 18.7\%  & 7,650  & 18.5\%  \\
            & Unknown & 2,030  & 14.7\%  & 1,925  & 13.9\%  & 2,038  & 14.7\%  & 5,993  & 14.4\%  \\
            & Error   & 0      & 0.0\%   & 0      & 0.0\%   & 0      & 0.0\%   & 0      & 0.0\%   \\
        \cline{2-10}
        \multicolumn{2}{|l|}{\textbf{Subtotal}} & 13,824 & 100.0\% & 13,824 & 100.0\% & 13,824 & 100.0\% & 41,472 & 100.0\% \\
        \hline
        \multirow{6}{*}{DeepSeek-R1}
            & Act     & 468    & 3.4\%   & 460    & 3.3\%   & 505    & 3.7\%   & 1,433  & 3.5\%   \\
            & Attend  & 2,964  & 21.4\%  & 3,402  & 24.6\%  & 3,722  & 26.9\%  & 10,088 & 24.3\%  \\
            & Track*  & 842    & 6.1\%   & 792    & 5.7\%   & 797    & 5.8\%   & 2,431  & 5.9\%   \\
            & Track   & 3,077  & 22.3\%  & 2,609  & 18.9\%  & 2,411  & 17.4\%  & 8,097  & 19.5\%  \\
            & Unknown & 6,469  & 46.8\%  & 6,558  & 47.4\%  & 6,388  & 46.2\%  & 19,415 & 46.8\%  \\
            & Error   & 4      & 0.0\%   & 3      & 0.0\%   & 1      & 0.0\%   & 8      & 0.0\%   \\
        \cline{2-10}
        \multicolumn{2}{|l|}{\textbf{Subtotal}} & 13,824 & 100.0\% & 13,824 & 100.0\% & 13,824 & 100.0\% & 41,472 & 100.0\% \\
        \hline
        \multirow{6}{*}{Gemini Flash 1.5}
            & Act     & 456    & 3.3\%   & 870    & 6.3\%   & 848    & 6.1\%   & 2,174  & 5.2\%   \\
            & Attend  & 1,270  & 9.2\%   & 1,602  & 11.6\%  & 1,748  & 12.6\%  & 4,620  & 11.1\%  \\
            & Track*  & 1,087  & 7.9\%   & 945    & 6.8\%   & 834    & 6.0\%   & 2,866  & 6.9\%   \\
            & Track   & 8,944  & 64.7\%  & 8,495  & 61.5\%  & 8,239  & 59.6\%  & 25,678 & 62.0\%  \\
            & Unknown & 2,067  & 15.0\%  & 1,912  & 13.8\%  & 2,155  & 15.6\%  & 6,134  & 14.8\%  \\
            & Error   & 0      & 0.0\%   & 0      & 0.0\%   & 0      & 0.0\%   & 0      & 0.0\%   \\
        \cline{2-10}
        \multicolumn{2}{|l|}{\textbf{Subtotal}} & 13,824 & 100.0\% & 13,824 & 100.0\% & 13,824 & 100.0\% & 41,472 & 100.0\% \\
        \hline
        \multirow{6}{*}{ChatGPT 4o-mini}
            & Act     & 5,852  & 42.3\%  & 6,121  & 44.3\%  & 5,872  & 42.5\%  & 17,845 & 43.0\%  \\
            & Attend  & 286    & 2.1\%   & 416    & 3.0\%   & 516    & 3.7\%   & 1,218  & 3.0\%   \\
            & Track*  & 54     & 0.4\%   & 38     & 0.3\%   & 95     & 0.7\%   & 187    & 0.5\%   \\
            & Track   & 4,388  & 31.7\%  & 3,937  & 28.5\%  & 4,066  & 29.4\%  & 12,391 & 29.9\%  \\
            & Unknown & 3,244  & 23.5\%  & 3,312  & 24.0\%  & 3,275  & 23.7\%  & 9,831  & 23.7\%  \\
            & Error   & 0      & 0.0\%   & 0      & 0.0\%   & 0      & 0.0\%   & 0      & 0.0\%   \\
        \cline{2-10}
        \multicolumn{2}{|l|}{\textbf{Subtotal}} & 13,824 & 100.0\% & 13,824 & 100.0\% & 13,824 & 100.0\% & 41,472 & 100.0\% \\
        \hline
        \multicolumn{2}{|l|}{\textbf{Total}} 
            & \textbf{69,120} &  & \textbf{69,120} &  & \textbf{69,120} &  & \textbf{207,360} & \\
        \hline
    \end{tabular}
    \caption{Vulnrichment ground truth and LLM SDO Distributions across MWSs (cf. \S\ref{subsec:rq2_discussion}). }
\end{table}

\end{document}

%% file: files/intro.tex
Effective vulnerability management is critical for organizations aiming to mitigate cybersecurity threats and minimize risk exposure. A core challenge within this process is \textit{vulnerability prioritization}--the task of determining which vulnerabilities require immediate attention and which can be deferred. Traditionally, this task has relied on manual or semi-automated processes within the broader vulnerability management lifecycle. These methods often require specialized tools and expert judgment to interpret alerts and assess potential impacts, making the process both resource-intensive and time-consuming~\cite{crowdstrike2024appsec}. Under high-pressure conditions, this burden can lead to misclassifications, delayed responses, or missed critical threats.

To address these limitations, researchers have turned to machine learning and deep learning techniques~\cite{fu2024aibughunter, elder2024survey, noauthor_software_2024, jisoo2024research}. These models use historical data to classify and prioritize vulnerabilities based on statistical patterns. However, three key limitations remain: (1) a scarcity of large-scale, high-quality labeled datasets~\cite{alzubaidi2023survey}; (2) high dependency on the distribution and completeness of training data, which hinders generalization to novel vulnerabilities~\cite{guo2024comprehensive}; and (3) difficulty adapting to evolving threats, including zero-day exploits and emerging attack vectors~\cite{kan2024tesseract}.

Recently, LLMs like ChatGPT, Claude, Gemini, and DeepSeek have emerged as promising tools for automating vulnerability analysis. Pre-trained on vast textual corpora, these models can interpret unstructured and semi-structured data with minimal reliance on manual labeling~\cite{zhao2023survey}. LLMs have shown strong performance in related security domains, including source code analysis~\cite{xu2022systematic} and threat intelligence classification~\cite{CHEN2024104016}, suggesting their potential for vulnerability prioritization.

In this paper, we present the first systematic evaluation of four state-of-the-art LLMs--ChatGPT 4o-mini (ChatGPT), Claude 3 Haiku (Claude), Gemini Flash 1.5 (Gemini), and DeepSeek R1 (DeepSeek)--across twelve PTs for automated vulnerability prioritization. We ground our analysis in the {Stakeholder-Specific Vulnerability Categorization} (SSVC) framework~\cite{noauthor_ssvc_nodate-1, keizman_ssvc_2024, noauthor_cisa_nodate}, where analysts assess the four key SSVC decision points (SDPs): {Exploitation}, {Automation}, {Technical Impact}, and {Mission \& Wellbeing}, leading to prioritized SSVC decision outcomes (SDOs): {Track}, {Track*}, {Attend}, or {Act} (cf. Table~\ref{tab:ssvc_outcomes}). 

Despite their potential, LLMs have not been systematically evaluated for their utility in supporting security analysts in context-aware vulnerability prioritization. To fill this gap, our study investigates the effectiveness of LLMs across 12 PTs in classifying SDPs. Specifically, we aim to answer the following research questions:



\begin{itemize}
 \item \textbf{$\mathbf{RQ1}$}: Which LLM demonstrates the highest accuracy in analyzing vulnerability data to recommend SDP responses?
 \item 
 \textbf{$\mathbf{RQ2}$}: How accurately do LLM-generated SDPs, when processed through the SSVC decision tree, lead to correct SDOs? 
\item \textbf{$\mathbf{RQ3}$}: To what extent does the choice of PT influence the performance of LLMs in vulnerability analysis tasks? 
\item \textbf{$\mathbf{RQ4}$}: For each SDP, which LLM and PT combination results in the best predictive performance? 
\item \textbf{$\mathbf{RQ5}$}: Do LLMs suffer from recommending false positive or false negative SDOs?
\end{itemize}



To address these research questions, we developed a crawling and analysis framework (\S~\ref{sec:estudy}) that presents the first systematic evaluation of four state-of-the-art LLMs across twelve PTs for automated vulnerability prioritization. Our evaluation leverages a sample of 384 real-world vulnerabilities, curated from VulZoo~\cite{10.1145/3691620.3695345} and annotated in Vulnrichment~\cite{baran_cisa_2024}, which constitutes the ground truth dataset for this study. In total, we executed over 165,000 LLM queries and assessed model performance using F1-scores and Cohen's Kappa coefficient to measure agreement with the SSVC ground truth.

Our key contributions are as follows:
\begin{itemize}
    \item \textbf{Comprehensive benchmarking and comparative analysis.} We conducted a large-scale empirical study evaluating four LLMs across 12 PTs on 384 real-world vulnerabilities (\S~\ref{sec:pereval}). Gemini achieved the highest F1-score of 0.79 for \textit{Exploitation}, and was the only model to generate a majority of correct SDOs.
    \item {\bf Evaluation of PTs.} We found that exemplar-based PTs, such as Chain-of-Thought (CoT) and Few-shot with Explanation (FSWE), yielded the best results, with FSWE reaching an F1-score of 0.75 (\S~\ref{subsec:rq3_discussion}). 
    \item \textbf{Analysis of failure cases in LLM predictions.} We examined the potential of false positive and false negative limitations in LLM-driven vulnerability prioritization and emphasized the need for security analysts to remain critical in decision-making oversight (\S~\ref{subsec:rq5_discussion}). We found that all analyzed LLMs generated more false positives than false negatives SDOs, with Claude (8.9K), DeepSeek (26.6K), and ChatGPT (6.3K) producing more false positives than correct responses. Only Gemini (22.2K true positives vs. 11.9K false positives) achieved a majority of correct SDO predictions.

    
\end{itemize}

Our study highlights the current strengths and limitations of LLMs in operationalizing SSVC-based vulnerability prioritization.  We provide actionable insights on selecting LLM–PT pairings for different SDPs. While these models are not yet reliable enough to replace human analysts, they can augment decision-making pipelines when configured appropriately.

%% file: files/bg-rwork.tex
\label{subsec:vulcatfmwork}



Vulnerability prioritization frameworks provide structured methodologies to assess and prioritize security vulnerabilities based on their severity, potential impact, and likelihood of exploitation. These frameworks enable organizations to make informed, risk-based remediation decisions and to optimize the allocation of security analyst resources~\cite{spring2021prioritizing}. 

This section elaborates on three widely adopted vulnerability prioritization frameworks that differ in scope, decision support, and contextual awareness.

\subsection{Common Vulnerability Scoring System (CVSS)} 

The CVSS is a framework for describing the traits and impacts of vulnerabilities~\cite{mell_common_2007}. The CVSS comprises three metric groups: (1) Base, reflecting invariable traits of the vulnerability; (2) Temporal, reflecting traits that may change over time but not user environments; and (3) Environmental, reflecting traits specific to the user environment~\cite{mell_common_2007}. While it is a detailed system for capturing key data for a given vulnerability, the CVSS does not recommend a tailored action for security analysts to take in response to a vulnerability. 

The CVSS produces a severity score ranging from 0.1 to 10.0. However, it does not prescribe tailored remediation actions. Security analysts must often rely on static thresholds (e.g., scores above 7.0) to determine the urgency of a response. This arbitrary cutoff can be problematic when multiple vulnerabilities exceed the threshold, limiting the precision and effectiveness of prioritization strategies~\cite{HowShoul49:online}.


\subsection{Exploit Prediction Scoring System (EPSS)} 
The EPSS attempts to forecast the likelihood that a given vulnerability will be exploited within the next 30 days~\cite{TheEPSSM8:online}. Given that only 2\% - 7\% of vulnerabilities are exploited, the goal is to best allocate response efforts to those vulnerabilities most likely to be exploited~\cite{WhatisEP89:online}. Over 1000 data points are used to calculate an EPSS score via a machine learning model, including vulnerability datasets, social media activity, and offensive security tool developments~\cite{jacobs2023enhancing}. 

Although EPSS provides valuable predictive insights into exploitation likelihood, it lacks contextual awareness regarding the business impact of a vulnerability. Consequently, while it informs about the urgency of a potential exploit, it does not assist security analysts in understanding the operational significance or potential harm of the vulnerability to their specific environment~\cite{HowShoul49:online}.


\subsection{Stakeholder-Specific Vulnerability Categorization Framework} 
\label{subsec:ssvc}
To provide tailored vulnerability prioritization recommendations, the SSVC framework was developed by CISA and Carnegie Mellon University~\cite{noauthor_stakeholder-specific_nodate}. The SSVC enables organizations to move beyond generic severity scores by guiding analysts through a structured decision-making process that accounts for operational context, risk tolerance, and mission-critical assets.

At the core of the SSVC are the four SDPs, which security analysts assess to classify the severity and urgency of a vulnerability:

\begin{enumerate}
    \item \textbf{Exploitation:} Assesses whether the vulnerability is currently being exploited or has credible reports of exploitation in the wild.
    \item \textbf{Automatable:} Determines the extent to which the exploitation process can be automated, which may impact how quickly the threat can spread or be weaponized.
    \item \textbf{Technical Impact:} Evaluates the potential damage the vulnerability could cause, such as loss of confidentiality, integrity, or availability of critical systems.
    \item \textbf{Mission \& Wellbeing:} Reflects the potential harm to the organization's mission or human wellbeing if the vulnerability is exploited, emphasizing contextual impact beyond technical considerations.

\end{enumerate}


These SDPs are processed through a decision tree to yield one of four SDOs (further explained in Table~\ref{tab:ssvc_outcomes} in Appendix~\ref{app:sdo}):

\begin{enumerate}
    \item \textbf{Track:} Monitor the vulnerability but defer action unless the context changes.
    \item \textbf{Track*:} Monitor more closely; reassessment may be needed soon.
    \item \textbf{Attend:} Take some action, such as deploying mitigations or reviewing exposure.
    \item \textbf{Act:} Immediate remediation is required, typically via patching or isolation.

\end{enumerate}


In practice, security analysts use the SSVC to triage vulnerabilities in line with their organizational risk posture. For example, a vulnerability with active exploitation, high technical impact, and significant mission implications in a critical infrastructure setting (e.g., a power plant) would likely result in an \texttt{Act} decision, prompting immediate remediation efforts.

Because the SSVC relies on manual classification of the SDPs, it offers fine-grained and context-aware prioritization, though at the cost of scalability, especially for organizations managing thousands of vulnerabilities~\cite{TheSSVCm81:online}. Despite this limitation, the SSVC complements traditional scoring systems by delivering structured and actionable guidance~\cite{noauthor_we_nodate}.

{\bf The need for automating SSVC with LLMs. } 
Given the labor-intensive nature of SSVC assessments, automating this process using LLMs presents a compelling opportunity. To illustrate this, we applied the SSVC framework to \texttt{CVE-2018-0151}, a remote code execution vulnerability with confirmed exploitation. We simulated a scenario within a critical infrastructure environment, assigning a \textit{High} value to the Mission \& Wellbeing SDP due to potential harm to human life.

Vulnerability data were sourced from VulZoo, encompassing both semi-structured and unstructured information, and formatted according to prompt engineering recommendations. Using CoT prompting and contextual exemplars, we queried Gemini~\cite{google_gemini_flash_2024} through the OpenRouter API~\cite{noauthormodelsnodate}. To account for LLM output variability, we conducted three independent trials per prompt.

We then evaluated Gemini's performance using F1-scores across all four SDPs. For the Exploitation SDP, Gemini achieved a harmonic mean F1-score of 0.79, demonstrating strong consistency and predictive capability. This shows that, with well-structured prompts, LLMs can effectively automate SSVC decision-making, significantly reducing analyst workload and improving the speed and precision of vulnerability triage. Further methodological details are provided in \S~\ref{sec:estudy}.

\subsection{Related Work}
\label{subsec:vuldetection}

In this section, we present a summary (cf. Table \ref{tab:relatedwork}) of the literature in the field of artificial intelligence for vulnerability management. 

\begin{table*}[h!]
\centering

\begin{tabular}{p{2.5cm}p{5cm}p{4.5cm}p{3.5cm}}
\toprule
\textbf{Related Work} & \textbf{Vulnerability Prioritization Framework} & \textbf{Considers Organizational Context} & \textbf{Uses Off-the-Shelf LLMs} \\
\midrule
Ghaffarian et al. \cite{ghaffarian2017software} & General ML categorization & No & No \\
Harer et al. \cite{harer2018automated} & ML-based vulnerability detection & No & No \\
Tanga et al. \cite{tangaadvanced} & Anomaly detection & No & No \\
Hore et al. \cite{hore2023deep} & ML for dynamic prioritization & Some (dynamic nature) & No \\
CVSS-BERT \cite{shahid2021cvss} & CVSS vector mapping & No & No \\
Yin et al. \cite{YIN2020106529} & Exploitation likelihood prediction & No & No \\
Oniagbi \cite{oniagbi_evaluation_2024} & SIEM event classification & Some (SIEM data) & Yes \\
LocalIntel \cite{mitralocalintel2024} & Tailored vulnerability assessments & Yes & Yes (one LLM) \\
\midrule
\textbf{Our Work} & SSVC & Yes & Yes (multiple LLMs) \\
\bottomrule
\end{tabular}
\caption{Summary and comparison of related work.}
\label{tab:relatedwork}
\vspace{-0.45cm}
\end{table*}

{\bf Machine Learning for Vulnerability Analysis and Prioritization.} 
Security researchers have extensively leveraged machine learning and deep learning models to perform vulnerability analysis and prioritization. Ghaffarian et al. \cite{ghaffarian2017software} categorize these approaches into three main types: (1) Supervised Learning, where models are trained to infer specific functions by pairing inputs with desired outputs; (2) Unsupervised Learning, where models infer relationships within datasets without predefined labels; and (3) Reinforcement Learning, where models learn and adjust behavior based on reward functions to achieve desired outcomes. For instance, Harer et al. \cite{harer2018automated} demonstrated that machine learning models could effectively detect vulnerabilities in C and C++ code using static analysis results as ground truth. Similarly, Tanga et al. \cite{tangaadvanced} found that unsupervised learning models were effective in discriminating between benign and malicious network traffic, highlighting their potential for detecting novel and zero-day vulnerabilities due to their independence from predefined ground truth datasets. Hore et al. \cite{hore2023deep} developed Deep VULMAN, a reinforcement learning-based framework for dynamically prioritizing security threats and allocating resources, addressing the limitation of prior research that treated vulnerability triaging as a one-time decision process, which is suboptimal in the dynamic real-world context of vulnerabilities.

{\bf Vulnerability categorization with LLMs.} 
The use of LLMs for evaluating software vulnerabilities and suspicious network or system activity is an emerging area of research. One line of work focuses on Bidirectional Encoder Representations from Transformers (BERT), where content before and after a particular text is processed to derive contextual awareness, making it suitable at recognizing the syntax and semantics of language~\cite{HAUROGNE2024100598}. An example of this is CVSS-BERT~\cite{shahid2021cvss}, which maps vulnerability descriptions to CVSS vector string values, achieving a strong F1-Score performance. Yin et al. \cite{YIN2020106529} introduced ExBERT, a model capable of determining the likelihood of vulnerability exploitation from descriptions with high accuracy. Another line of research explores Generative Pre-trained Transformers (GPT), where next-word prediction functionality for a sequence of text allows for inferences to be developed between key terms and concepts~\cite{yenduri2023generative}. Oniagbi \cite{oniagbi_evaluation_2024} demonstrated the utility of LLMs in analyzing Security Information and Event Management (SIEM) data and classifying events as ``Interesting'' or ``Not Interesting''. Shaswata et al. \cite{mitralocalintel2024} developed LocalIntel, a GPT-3.5-turbo-based framework that combines global vulnerability knowledge with local organizational context to provide tailored vulnerability assessments.

{\bf Differentiation of our work.} 
Table \ref{tab:relatedwork} highlights the research gap and differentiates our work from existing literature. As discussed, while prior studies demonstrate the potential of LLMs for vulnerability detection, they generally do not employ vulnerability prioritization frameworks that explicitly account for individual organizational contexts, such as the SSVC. Furthermore, although custom LLMs exhibit promise, organizations may lack the requisite resources to train, host, and maintain such models \cite{urlana-etal-2025-size}, thus favoring readily available, off-the-shelf solutions. This disparity underscores the need for research that leverages off-the-shelf LLMs to deliver actionable, context-aware vulnerability prioritization recommendations.

%% file: files/estudy.tex
\begin{figure*}[th!]
    \centering
    \includegraphics[width=0.9\textwidth]{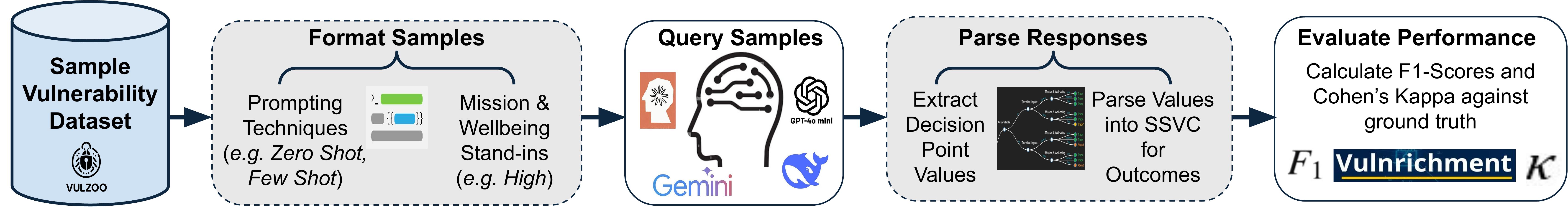}
    \caption{Overview of the vulnerability data collection, processing, prompting, and evaluation pipeline. 384 vulnerabilities were selected that were common to both the VulZoo and Vulnrichment datasets. Samples were parsed with the PTs and MWSs, and queried against the LLMs across three trials. Performances were evaluated by: (1) F1-Scores, comparing LLM SDP responses against Vulnrichment SDP values, with a harmonic mean calculated across the trials for each LLM-PT-SDP combination; and (2) Cohen's Kappa, comparing SDOs generated by parsing LLM SDPs and Vulnrichment SDPs through the SSVC decision tree for each MWS.}
    \label{fig:pipline}
    \vspace{-0.45cm}
\end{figure*}

Our study seeks to broadly explore the capability of the mainstream LLMs for vulnerability prioritization. Figure~\ref{fig:pipline} explains our data collection and analysis pipeline. 


\subsection{Vulnerability Datasets and Processing Pipeline}

\subsubsection{Vulnrichment: Annotated SSVC Dataset}
\label{subsec:vulnrichment}

The \textit{Vulnrichment} dataset~\cite{noauthorcisagovvulnrichment2024} is an annotated repository of real-world software vulnerabilities maintained by the U.S. Cybersecurity and Infrastructure Security Agency (CISA). As of November 22, 2024, Vulnrichment contained 45,621 curated vulnerability records, each indexed by a unique CVE identifier. The dataset is specifically designed to support SSVC-based risk assessment workflows, offering security analysts a pre-annotated subset of CVEs with fields aligned to the SSVC decision-making model.

Each record in the dataset includes several key SSVC-relevant attributes:
\begin{itemize}
    \item \textbf{CVE ID}: A globally unique identifier from the MITRE CVE program, used to reference specific vulnerabilities.
    \item \textbf{SSVC Technical Impact}: A qualitative assessment describing the severity of the vulnerability’s impact on confidentiality, integrity, or availability.
    \item \textbf{SSVC Exploitation}: A binary or categorical indicator denoting whether the vulnerability is known to be actively exploited in the wild.
    \item \textbf{SSVC Automatable}: A classification of whether the exploitation process can be automated, which informs propagation risk and exploitation scale.
\end{itemize}

Given the alignment between these fields and the SSVC framework, Vulnrichment serves as a suitable, albeit partial, \textit{ground truth dataset} for training and evaluating SSVC decision point (SDPs) predictions. However, it does not include values for the \textit{Mission \& Wellbeing} (M\&W) decision point, which is inherently stakeholder-specific and context-dependent.

\subsubsection{Handling Missing Mission \& Wellbeing Data}
\label{subsubsec:missing_mw}

The absence of stakeholder-specific M\&W values in Vulnrichment poses a challenge for complete SSVC evaluation. To approximate real-world usage, we introduced \textit{Mission \& Wellbeing Stand-ins} (MWSs) to simulate different organizational risk profiles. This approach is consistent with prior SSVC use cases~\cite{noauthor_we_nodate}, where analysts define M\&W levels based on organizational mission criticality.

We designed three MWSs representing common organizational risk scenarios (further detailed in Appendix~\ref{app:mission_wellbeing_standins}): (1) {\it High-risk scenario}: Critical infrastructure settings (e.g., power plants), where exploitation could endanger human life or national security.
    (2) {\it Medium-risk scenario}: Organizations with high business continuity demands (e.g., supermarkets), where service disruption would cause significant economic impact.
    (3) {\it Low-risk scenario}: Low-priority environments (e.g., leisure centers), where impact is minimal or isolated.


Each MWS was embedded into the system prompt during LLM queries to guide contextual reasoning. This enabled LLMs to simulate tailored M\&W decisions based on organizational characteristics. For example, in a power plant context, an LLM should reasonably infer a \textit{High} M\&W classification, reflecting elevated safety and operational risks. This strategy enabled consistent simulation of real-world analyst judgment without requiring inaccessible proprietary data.

\subsection{VulZoo Vulnerability Corpus}
\label{subsec:vulzoo}


To supplement Vulnrichment and broaden the range of LLM-evaluable vulnerability descriptions, we incorporated the \textit{VulZoo} dataset~\cite{10.1145/3691620.3695345}. VulZoo is a large-scale, open-source vulnerability corpus that aggregates data from 17 independent sources. It includes a comprehensive collection of 323,839 CVEs, enriched with structured, semi-structured, and unstructured metadata fields. These include technical descriptions, exploit details, publication references, affected product data, and severity indicators~\cite{VulZoopr8:online}.

Given the strengths of LLMs in interpreting semi-structured and unstructured data, VulZoo is particularly well-suited for evaluating LLM performance under realistic input conditions. For each CVE in our analysis, we extracted relevant fields—such as vulnerability summary, exploit status, and impact statements—and formatted them into prompt templates designed to elicit predictions for each SDP. Prompts were structured using twelve PTs including Zero-shot (ZS), Few-shot (FS), and CoT techniques.



\subsection{Selecting Sample Dataset} 
Processing all 45,000+ records in Vulnrichment—across multiple LLMs, PTs, MWSs, and trials—would be computationally and financially infeasible. To address this challenge while preserving statistical validity, we adopted a stratified random sampling approach.

Based on a 95\% confidence interval~\cite{SampleSi14:online}, 5\% margin of error, and an estimated response proportion of 50\%, we computed a required minimum sample size of 380 vulnerabilities. We ultimately selected a slightly larger sample of 384 CVE IDs (cf. Table~\ref{tab:experimental_parameters}) to ensure balanced representation across different risk scenarios and to align with related ongoing research. This sampling approach ensured that our evaluation was both rigorous and practical, while maintaining generalizability across the broader vulnerability population.


\begin{table}[h]
    \centering
    \begin{adjustbox}{width=\columnwidth}
        \begin{tabular}{p{5cm}|p{3cm}}
        \toprule
            
            \textbf{Parameter} & \textbf{Value} \\
            \midrule
            \hline
            \textbf{Sample Size} & 384 \\
            \hline
            \textbf{Number of PTs} & 12 \\
            \hline
            \textbf{Number of LLMs} & 4 \\
            \hline
            \textbf{Number of Trials} & 3 \\
            \hline
            \textbf{Number of MWSs} & 3 \\
            \hline
            \textbf{Total Queries} & \textbf{165,888} \\
            \hline
            \bottomrule
        \end{tabular}
    \end{adjustbox}
    \caption{Total number of queries sent to OpenRouter.}
    \label{tab:experimental_parameters}
    \vspace{-0.45cm}
\end{table}

\subsection{Parsing and Prompting Sample Vulnerabilities}
\label{subsec:parse_sample}

\subsubsection{PTs}
\label{subsubsec:prompt_techniques}
To evaluate LLMs under diverse interaction paradigms, we utilize a set of PTs informed by the taxonomy proposed by Tony et al.~\cite{tony_prompting_2024}. PTs are grouped into five major categories:

\begin{enumerate}
    \item \textbf{Root Techniques:} Basic, standalone prompts used as a baseline for LLM performance (e.g., ZS, OS).
    \item \textbf{Refinement-Based Techniques:} Iterative prompting techniques where model outputs are revised based on feedback.
    \item \textbf{Decomposition-Based Techniques:} Techniques that segment complex problems into smaller tasks to improve reasoning (e.g., step-by-step).
    \item \textbf{Reasoning-Based Techniques:} Prompts that encourage logical inference, such as CoT.
    \item \textbf{Priming Techniques:} Techniques that include cultural memes to orient the model towards desirable outputs (e.g., role-play).
\end{enumerate}

Some PTs--such as \textit{Progressive Hint} and \textit{Least-to-Most}--require users to provide feedback or structure sub-tasks interactively across multiple prompts~\cite{zhou_least--most_2023}. While these methods have shown promise in guiding LLMs toward more accurate outputs, they place a significant burden on the user to act as a facilitator of the model’s reasoning process.
To reduce cognitive overhead for analysts working in high-pressure environments, we excluded Progressive Hint and Least-to-Most PTs from our study. Instead, we focused on single-shot prompt executions that could feasibly be scaled and integrated into automated pipelines. The full list of included PTs is provided in Appendix~\ref{app:prompt_categories_techniques}.


\subsubsection{Prompt Engineering and Best Practice Formatting}
\label{subsubsec:best_practice_formatting}
Research has shown that the formulation of prompts can significantly influence LLM performance~\cite{lee2024optimize}. To optimize output quality, we incorporated several best practices from prompt engineering literature:
(1) {\it Imperative Tone--}commands framed in imperative form (e.g., ``Return the SSVC decision point...'') lead to more consistent and accurate outputs. 
    (2) {\it Positive Phrasing--}avoiding negations improves clarity and model confidence~\cite{huang_utilizing_2024}.
    (3) {\it Acknowledgement of Uncertainty--}prompts encouraged LLMs to respond with \texttt{Unknown} when insufficient information was present, which mitigates overconfident hallucinations.

In addition, we adopted a structured prompt format using XML-style tags to clearly delineate instructions, examples, context, and input data. Prior work~\cite{noauthor_use_nodate, noauthor_openai_nodate, noauthor_structure_nodate} has shown that structured inputs help LLMs distinguish between different prompt segments, leading to improved parsing and interpretability. This tagging schema was consistently applied across all PTs to reduce variability in model responses and enhance reproducibility.


\subsection{Selecting, Configuring, and Querying LLMs}
\label{subsec:select_llms}

\subsubsection{Model Selection Criteria}
To ensure cost-effective and practically deployable evaluation, we selected LLMs that demonstrated strong performance on public benchmarks while maintaining relatively low token cost. Table~\ref{tab:llm_specs} lists the models studied in this work. The models include: (1) {ChatGPT 4o-mini}, (2) {Claude 3 Haiku}, (3) {Gemini Flash 1.5}, and (4) {DeepSeek R1}. These models are chosen based on their availability on widely used cloud platforms--Amazon Web Services (AWS), Microsoft Azure, and Google Cloud--which collectively represent over 66\% of the global cloud computing market~\cite{Kingsley2024}. This criterion ensures ease of integration into enterprise IT infrastructures and aligns with the operational requirements of many organizations.

Due to prohibitive cost constraints, we excluded the premium-tier models, such as ChatGPT-4o, Claude Opus 3, and Gemini Pro 1.5. 
At the time of writing, these models range from \$6.25 to \$90 per million tokens~\cite{noauthormodelsnodate}, making them impractical for large-scale academic experimentation.

During this study, DeepSeek released its R1 model~\cite{guo2025deepseek}, which demonstrated performance comparable to premium-tier models while maintaining pricing similar to mid-range alternatives. This prompted its inclusion as a representative case to assess whether newer cost-efficient models could bridge the performance gap, offering enterprise-ready capabilities without the financial burden of premium APIs.



\begin{figure*}[h]
    \centering
    \includegraphics[width=0.9\textwidth]{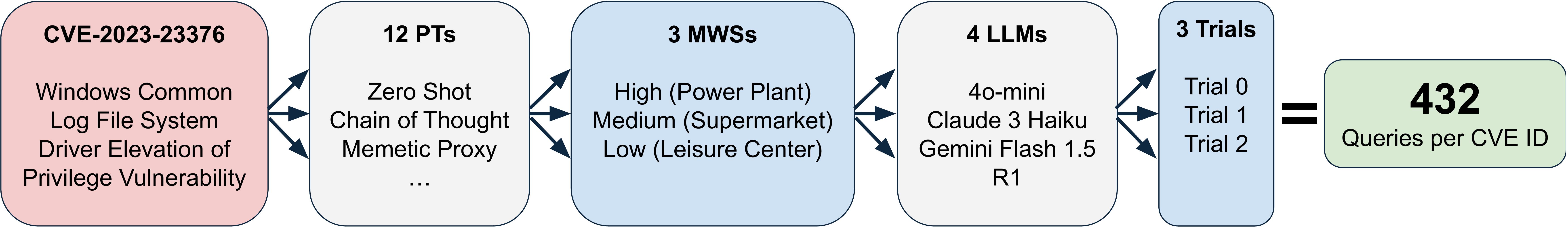}
    \caption{Calculation of number of queries per CVE. When multiplied by 384 CVEs, with 12 PTs, and 3 different MWSs, we queried 4 distinct LLMs with 165,888 queries via OpenRouter API.}
    \label{fig:cve-query-calculation}
    \vspace{-0.35cm}
\end{figure*}

\begin{table}[h]
    \centering
    \begin{adjustbox}{width=\columnwidth}
        \begin{tabular}{l|r|r|r|r}
        \toprule
        \hline            
            \textbf{Model} & \textbf{Window} & \textbf{Input} & \textbf{Output} & \textbf{Total} \\
            & \textbf{(tokens)} & \textbf{(\$/M)} & \textbf{(\$/M)} & \textbf{(\$/M)} \\
            \hline
            \textbf{ChatGPT 4o-mini} & 128K & 0.15 & 0.60 & 0.75 \\
            \hline
            \textbf{Claude 3 Haiku} & 200K & 0.25 & 1.25 & 1.50 \\
            \hline
            \textbf{Gemini Flash 1.5} & 1M & 0.08 & 0.30 & 0.38 \\
            \hline
            \textbf{DeepSeek R1} & 164K & 0.55 & 2.19 & 2.75 \\
            \hline
            \bottomrule
        \end{tabular}
    \end{adjustbox}
    \caption{LLM context windows and pricing, as per ~\cite{noauthormodelsnodate}.}
    \label{tab:llm_specs}
    \vspace{-0.45cm}
\end{table}
\subsubsection{LLM Query Strategy}
All models were queried using the twelve selected PTs and evaluated across three independent trials to account for output variability due to model nondeterminism. Each query presented the model with a vulnerability record formatted using best-practice prompting, and responses were evaluated for their accuracy in classifying the four SDPs. Further details of our LLM evaluation pipeline are provided in \S~\ref{sec:pereval} and \S~\ref{sec:perevalresults}.

\subsection{Configuring and Querying LLMs}
\label{subsec:llm_config_query}
\subsubsection{LLM Configuration Parameters}
To ensure a controlled and reproducible experimental environment, we accessed all selected LLMs via the OpenRouter API~\cite{noauthormodelsnodate}, a unified gateway that facilitates access to multiple proprietary language models. The API allows customization of various decoding parameters, which influence the randomness and diversity of model-generated responses. In our experiments, we focused on the following key parameters: (1) {\it Temperature--}controls the level of randomness in token selection. A temperature of 0 results in deterministic outputs, whereas a value closer to 1 encourages more creative and varied responses~\cite{nick_2_nodate}. We set this to 0.7 to balance coherence and diversity. (2) {Top-P (Nucleus Sampling)--}instructs the LLM to consider tokens whose cumulative probability mass does not exceed a specified threshold. Lower values (e.g., 0.1) lead to more focused and conservative outputs, while higher values (approaching 1) increase variability~\cite{noauthor_llm_nodate-1}. We used a value of 0.7 to allow plausible yet contextually consistent predictions. (3) {Top-K Sampling--}restricts the model to selecting from the top $K$ highest-probability tokens at each step. Higher values increase diversity, while lower values yield more deterministic behavior~\cite{noauthor_ibm_2024}. We configured this parameter to 50, in line with recent research showing that this range offers a good trade-off between novelty and fidelity~\cite{Complete15:online, Seo2025}.

This parameter configuration (Temperature = 0.7, Top-K = 50, Top-P = 0.7) is comparable to prior LLM benchmarking work as a balanced strategy to promote output variation without introducing excessive deviation from contextually appropriate outputs.

\subsubsection{Querying LLMs} Given that 
language models are inherently stochastic and can produce different outputs when queried with the same prompt multiple times. To account for this non-determinism and evaluate the reliability of model outputs, we implemented a multi-trial query strategy inspired by previous LLM evaluation protocols~\cite{atil_llm_2024}.


Each prompt was submitted to each LLM three times per CVE instance, across all combinations of PT and MWS. This yielded a robust distribution of outputs for each vulnerability and allowed us to evaluate consistency, accuracy, and reasoning stability. The total number of queries issued is summarized in Table~\ref{tab:experimental_parameters}, and the calculation methodology is visualized in Figure~\ref{fig:cve-query-calculation}.

To enhance transparency and facilitate traceability of LLM decision logic, we adopted a structured, two-stage approach to response generation. Instead of instructing the LLMs to directly output an SDO, we explicitly prompted them to return values for each of the four SDPs: \textit{Exploitation}, \textit{Automatable}, \textit{Technical Impact}, and \textit{Mission \& Wellbeing}.


We then applied the SSVC decision tree to the LLM-generated SDP values to derive the corresponding SDOs programmatically. This decoupled process serves several important purposes. \textit{Firstly}, it allows analysts to trace how the final decision was derived from individual risk dimensions. \textit{Secondly}, it prevents the model from shortcutting its reasoning based on heuristics or memorized outcome mappings. \textit{Thirdly}, it provides transparency for post hoc auditing and evaluation of model reliability across specific SDPs.

Our method of LLMs' configuration and querying mirrors real-world security workflows, where analysts often assess each risk dimension individually before deciding on a prioritization outcome. By mimicking this process, our methodology not only facilitates rigorous evaluation but also better simulates how LLMs may be deployed in practical vulnerability triage pipelines.

\subsection{Evaluating Performance}
\label{sec:pereval}
\subsubsection{Evaluating SDP Classification Accuracy}
\label{subsec:sdp_eval}
To assess how effectively each analyzed LLM and PT combination can classify SDPs, we evaluated model outputs against the Vulnrichment dataset~\cite{baran_cisa_2024} ground truth. Each LLM response was compared on a per-SDP basis: \textit{Exploitation}, \textit{Automatable}, \textit{Technical Impact}, and \textit{Mission \& Wellbeing}.

For this evaluation, we adopted the {F1-score} metric, widely used in classification tasks due to its ability to balance \textit{precision} and \textit{recall}. It is particularly valuable in contexts where class imbalance exists or where false positives and false negatives are of equal concern. The F1-score is formally defined as:


\begin{equation}
\text{F1-Score} = 2 \times \frac{(\text{precision} \times \text{recall})}{(\text{precision} + \text{recall})}
\label{eq:f1_score}
\end{equation}

Precision measures the proportion of correctly predicted positive instances among all predicted positive instances, while recall measures the proportion of correctly predicted positive instances among all actual positive instances. 
The F1-score provides a harmonic mean of precision and recall, penalizing extreme imbalances between the two. Based on machine learning best practices~\cite{noauthor_f-score_2023}, scores $\geq$ 0.8 are indicative of strong performance, while scores between 0.5 and 0.8 suggest moderate performance.

To ensure reliability across the inherent stochastic nature of LLM outputs, we conducted three independent trials for each LLM–PT–CVE combination. We then computed the \textit{harmonic mean} of the F1-scores across these trials. The harmonic mean is particularly sensitive to low outlier values and less influenced by high values~\cite{Komić2011}, making it an appropriate choice to identify consistently high or low-performing combinations. This methodology provides a robust comparative analysis of LLM capability in classifying SDPs, highlighting not only peak performance but also consistency across multiple runs.

\subsubsection{Evaluating Agreement on Final SDO Outcomes}
\label{subsec:sdo_eval}
To assess end-to-end decision-making accuracy, we extended our evaluation to include final SSVC stakeholder decision outcomes (SDOs). For each vulnerability, we parsed both the LLM-generated and ground-truth SDP values into the official SSVC decision tree, which deterministically maps SDP combinations to one of four outcomes: \textit{Track}, \textit{Track*}, \textit{Attend}, or \textit{Act}. We then calculated both {\it Weighted} and {\it Unweighted} Cohen's Kappa scores~\cite{Weighted73:online} to measure the level of agreement between the LLM-derived SDOs and the Vulnrichment ground-truth SDOs across the three Mission \& Wellbeing Stand-in (MWS) scenarios. 
{Unweighted Cohen's Kappa} measures the raw agreement between two categorical rating sets, correcting for chance agreement. While Weighted Cohen's Kappa incorporates the severity of disagreement, it assigns partial credit when predictions deviate slightly from the ground truth. A Weighted Cohen's Kappa is useful to identify the extent of the differences between the LLM SDOs and Vulnrichment SDOs. For example, if the Vulnrichment ground truth stipulates an \textit{Act} outcome, an LLM that recommends \textit{Attend} would perform better than an LLM that recommends \textit{Track}. 

This is particularly important in SSVC, where decisions have ordinal significance. For example, misclassifying an \textit{Act} outcome as \textit{Attend} is less critical than misclassifying it as \textit{Track}. Weighted Kappa allows us to quantify the magnitude of such errors~\cite{Weighted73:online}, offering a more nuanced evaluation of LLM decision-making quality.

Together, the SDP-level F1-score and SDO-level Cohen's Kappa metrics provide a comprehensive evaluation of how well each LLM–PT combination performs both at the fine-grained classification level and at the holistic prioritization level. This dual-layer evaluation supports our analysis of whether LLMs can meaningfully support or augment vulnerability triage workflows in realistic, high-stakes environments.

